\documentclass{JHEP3}
\usepackage{amsmath}
\usepackage{cite}
\input epsf
\usepackage{epsfig}
\usepackage{amssymb}
\usepackage[active]{srcltx}

\def \be {\begin{equation}}
\def \ee {\end{equation}}
\def \bea {\begin{eqnarray}}
\def \eea {\end{eqnarray}}
\def \nn {\nonumber}

\DeclareMathOperator{\arcsinh}{arcsinh}

\def\bd{\begin{document}}
\def\ed{\end{document}}
\def\nn{\nonumber}
\def\bea{\begin{eqnarray}}
\def\eea{\end{eqnarray}}
\let\bm=\bibitem
\let\la=\label

\def\Label#1{\label{#1}%
  \smash{\hbox to0pt{\raise1ex\hbox{\tiny[#1]}\hss}}}
\newcommand{\bbibitem}[1]{\bibitem{#1}\marginpar{#1}}

\title{On some 1/4 BPS Wilson-'t Hooft loops  }
\author{Chang-Yong Liu\footnote{email address:
liuchangyong@nwsuaf.edu.cn} \ \ and \ \ Li Qin
\\
College of Science, Northwest A\&F University, Yangling, Shaanxi
712100, China}
\date{\today}
\abstract{ In this paper, we investigate the 1/4 BPS Wilson-'t Hooft
loops in $\mathcal{N}$=4 supersymmetric Yang-Mills theory. We use
the bulk D3-brane solutions with both electric and magnetic charges
on its world-volume to  describe some of 1/4 BPS Wilson-'t Hooft
loops. The D3-brane supersymmetric solutions are derived form
requiring $\kappa$-symmetry. We find the two consistent constraints
for Killing spinors and calculate the conserved charges of straight
1/4 BPS Wilson-'t Hooft loops and expectation values of circular 1/4
BPS Wilson-'t Hooft loops separately.}

\begin{document}

\section{Introduction}

 Wilson loop operators are interesting observables in gauge field theories. These operators can be viewed as
 the worldlines of a very massive quark with the electric charge.
In $\mathcal{N}$=4 SYM, the Wilson loops have the general form \be
W(\mathcal{C})=Tr\mathcal{P}\exp(i\int_{\mathcal{C}}(A_{\mu}\dot{x}^{\mu}+\Phi_i\theta^i|\dot{x}|)ds),
\ee where $x^i$ is the parametrization of the loop $\mathcal{C}$ and
$\theta^i$ is a unit vector in $R^6$. The $A_{\mu}$ are the gauge
fields and $\Phi_i$ are the six scalars in the adjoint
representation.

  According to the $AdS/CFT$ correspondence, the expectation values of fundamental representation Wilson
  loops are calculated on the AdS side as fundamental string action bounded by the curve $\mathcal{C}$ at
the boundary \cite{Mal97, Gubser:1998bc, Witten:1998qj,
Aharony:1999ti,Rey:1998ik, Mal98, Gross}: \be
W(\mathcal{C})=\int_{\partial X=\mathcal{C}}\mathcal{D}X
\exp(-\sqrt{\lambda}S[X]), \ee where the $S[X]$ is the string
action. For large $\lambda$, the expectation value of the Wilson
loop is given by the area $A$ of the minimal surface bounded by
$\mathcal{C}$ as \be W(\mathcal{C})\sim \exp(-\sqrt{\lambda}A). \ee

These fundamental strings sweep out a worldsheet in AdS side. There are two simple 1/2
BPS Wilson loops: the infinitely straight line and circular Wilson loop.

  For the infinitely straight line Wilson loop, its expectation value is simple
\be W_{line}=1. \ee The infinitely straight line Wilson loop
preserves half of the supersymmetries. For the circular Wilson loop
which also preserves half the supersymmetries, its expectation value
is \be W_{circle}=\exp(\sqrt{\lambda}). \ee The circular Wilson loop
can be calculated in perturbation theory by reducing the calculation
of the rainbow/ladder diagrams to matrix model \cite{Zarembo} \be
W_{circular}=\frac{1}{Z}\int
\mathcal{D}M\frac{1}{N}Tre^Me^{-\frac{2N}{\lambda}TrM^2}. \ee The
leading behavior at large N, it is \be W_{circular}\sim
\frac{2}{\sqrt{\lambda}}I_1(\sqrt{\lambda})\sim e^{\sqrt{\lambda}},
\ee where $I_1(x)$ is the Bessel function.

 For Wilson loops with high rank representation of
gauge group, we must consider the interaction between strings. These
stings can blow up to D3-branes and/or D5-branes configurations
corresponding to symmetric and anti-symmetric representation
separately
\cite{Drukker,Yamaguchi:2006D5,Gomis:2006sb,Gomis:2006im,Drukker01,Akemann:2001st,Myers:1999,
Rodriguez2006,Okuyama:2006jc,Yamaguchi:2006te,Lunin:2006xr,D'Hoker:2007fq,Hartnoll:2006is,Faraggi:2014tna}
analogous to the gaint gravitons
\cite{Grisaru:2000zn,Hashimoto:2000zp,McGreevy:2000cw,Lin:2004nb}.
The expectation value of 1/2 BPS circular Wilson loop in the
symmetric representation described by D3-brane is \be \label{cj}
W_{sym}=e^{2N(\kappa\sqrt{1+\kappa^2}+\arcsinh\kappa)} \ee with
$\kappa=\frac{k\sqrt{\lambda}}{4N}$. The expectation value of BPS
circular Wilson loop in the Anti-symmetric representation described
by D5-brane is \be
W_{asym}=e^{\frac{2N\sqrt{\lambda}}{3\pi}\sin^3\theta_k}, \ee where
$\theta_k$ is related to $k$ by \be
k=\frac{2N}{\pi}(\frac{1}{2}\theta_k-\frac{1}{4}\sin 2\theta_k). \ee

  The $\mathcal{N}$=4 SYM and the type IIB string theory have the electric-magnetic duality $SL(2,Z)$.
   Under S-duality, the electric charge is replaced with the magnetic charge. We need
consider another important loops known as Wilson-'t Hooft loops. The
Wilson-'t Hooft loops are the dyon worldlines wich carry both the
electric and magnetic charges in gauge
theory\cite{Kapustin:2005py,Pucci:2012sx}. In purely electric case,
they reduce to the Wilson loops. They are classified by a pair of
weights (electric and magnetic) for the gauge group and its magnetic
dual, modulo the action of the Weyl group. From the string theory,
the Wilson-'t Hooft loops are the $(F1,D1)$ bound states ending on
the worldvolume of D3-branes\cite{Witten:1995im}. For a
$(n,m)$-string, the string tension is \be
\tau_{n,m}=\frac{\sqrt{n^2+m^2/g_s^2}}{2\pi\alpha'}, \ee which is
invariant under $S$-duality:
$(n,m,g_s,\alpha')\leftrightarrow(-m,n,g_s^{-1},\alpha'g_s)$. The
1/2 BPS Wilson-'t Hooft loops have been discussed in \cite{ChenHe}.
The expectation value of 1/2 circular BPS Wilson-'t Hooft loop is
the same as (\ref{cj}) with \be \kappa^2=
\frac{\pi}{4N}\frac{|n+m\tau|^2}{Im \tau}, \ee which is invariant
under the  $SL(2,Z)$ transformation \bea
\tau &\rightarrow& \frac{a\tau+b}{c\tau+d} \nn \\
ad-bc=1&,& a,b,c,d\in \mathbb{Z}, \nn
\eea

where the  Yang-Mills coupling
$\tau$ is related to the string coupling as following
\be
\tau=\frac{4\pi i}{g^2_{YM}}+\frac{\theta}{2\pi}=\frac{i}{g_s}+\frac{\chi}{2\pi}.
\ee

  In this paper, we study the 1/4 BPS Wilson-'t Hooft loops using D-branes description.
We consider 1/4 BPS Wilson-'t Hooft loops in symmetric
representation. Similar to the $F1$'s, these can be studed by using
the D3-branes configuration. The paper is organized as follows. In
section 2, we study the Wilson-'t Hooft loops with insertions using
D3-brane description and conserved charges. In section 3, we discuss
the 1/4 BPS circular Wilson-'t Hooft loops. We end with the
conclusions and discussions.

\section{Wilson-'t Hooft loops with insertions}
\subsection{Review of the straight 1/4 BPS Wilson loops}

  The 1/4 BPS Wilson loops can be constructed by Wilson loops with two insertions\cite{Drukker:2006xg}. These deformations of Wilson
  loop are related to a certain spin-chain systems and can be solved by
Bethe ansatz. Without the local insertions, these supersymmetric
Wilson loops preserve half the supersymmetries of the vacuum.
Considering 1/2 BPS local insertions, we can construct the the 1/4
BPS Wilson loops on $R\times S^3$ \bea
W_{Z^J}=TrP[Z^J(-\infty)e^{i\int_{-\infty}^{\infty}(A_t(t,0)+\Phi_3(t,0))dt}\overline{Z}^J(\infty)e^{i\int_{\infty}^{-\infty}(A_t(t,\pi)+\Phi_3(t,\pi))dt}].
\eea Where $Z=\Phi_1+i\Phi_2$ is a half-BPS complex scalar. The
angle $0$ and $\pi$ are two points on $S^3$.

  In the dual string theory on $AdS_5 \times S^5$, the 1/4 BPS Wilson loops in supergravity extend
to the two lines on the boundary which one line run up and another
line run down at antipodal point on the $R\times S^3$ boundary with
one of the insertions to the infinite past and the other one to
future infinity in global Lorentzian $AdS_5$.

  It is convenient to use the following global coordinates for $AdS_5$:
  \bea \label{metric}
  \frac{ds^2}{L^2}=-\cosh^2\rho dt^2+d\rho^2+\sinh^2\rho(d\chi^2+\sin^2\chi(d\vartheta^2+\sin^2\vartheta d\varphi^2))+d\theta^2+\sin^2\theta d\phi^2.
  \eea
Where $L^4=\lambda \alpha'^2$ is the radius of $AdS_5$ and $S^5$.
The Green-Schwarz string action is \be \mathcal{S}=\frac{L^2}{4\pi
\alpha'}\int d\sigma d\tau \sqrt{-h}h^{\alpha\beta}[-\cosh^2\rho
\partial_{\alpha}t\partial_{\beta}t+
\partial_{\alpha}\rho\partial_{\beta}\rho+\partial_{\alpha}\theta\partial_{\beta}\theta+\sin^2\theta \partial_{\alpha}\phi\partial_{\beta}\phi].
\ee Adopting a periodic ansatz \be \rho=\rho(\sigma),~~t=\omega
\tau,~~\theta=\theta(\sigma),~~\phi=\omega_1\tau, \ee the equations
of motion are \bea
\rho''-\omega^2\cosh\rho \sinh\rho=0, \nn \\
\theta''+\omega_1^2 \cos\theta \sin\theta=0. \nn \eea The solution
to the string equations of motion is \be \label{stingsolution}
\phi=t,~~\sin\theta=\frac{1}{\cosh\rho}. \ee This solution satisfies
the correct boundary condition. The surface approaches the boundary
of $AdS_5$ and get to the north-pole of $S^2$ associated to the
scalar $\Phi_3$ at $\sigma=0$. As $\sigma \rightarrow \infty$, the
string comes close to the center of $AdS_5$ and rotates around the
equator of $S^2$ carrying the angular momentum related to $Z^J$.
There are two parts to the sting: at $\chi=0$ and at $\chi=\pi$.
They are continuously connected to each other beyond $\rho=0$.

Using the supersymmetry analysis, the string solution preserves 1/4
of the supersymmetries. The $\kappa$-symmetry projector associated
with a fundamental string is \be \label{susys}
\Gamma=\frac{1}{\sqrt{-\det
g}}\partial_tx^{\mu}\partial_{\rho}x^{\nu}\gamma_{\mu}\gamma_{\nu}K.
\ee Where K acts on spinors by complex conjugation.
$\gamma_{\mu}=e^a_{\mu}\Gamma_a$ and $\Gamma_a$ are constant tangent
space gamma-matrices.
$\Gamma_{\star}=\Gamma^0\Gamma^1\Gamma^2\Gamma^3\Gamma^4$ is the
product of all gamma matrices in the $AdS_5$ direction. The number
of supersymmetries preserved by the string is the number of the
independent solutions to the equation $\Gamma\epsilon=\epsilon$. The
dependence of the Killing spinors on the relevant coordinates is
written as \be
\epsilon=e^{-\frac{i}{2}\rho\Gamma_{\star}\Gamma_1}e^{-\frac{i}{2}t\Gamma_{\star}\Gamma_0}e^{-\frac{i}{2}\theta\Gamma_{\star}\Gamma_5}
e^{\frac{1}{2}\phi\Gamma_{56}}\epsilon_0. \ee Where $\epsilon_0$ is
a constant chiral complex 16-component spinor. This satisfies the
Killing spinor equation \be
(\partial_{\mu}+\frac{1}{4}\omega^{ab}_{\mu}\Gamma_{ab}+\frac{i}{2L}\Gamma_{\star}\gamma_{\mu})\epsilon=0.
\ee Using the vielbeins \be e^0=L\cosh\rho dt, ~~e^1=L d\rho,~~
e^5=Ld\theta, ~~e^6=L\sin\theta d\phi \ee

and inserting the string solution (\ref{stingsolution}) into the expression (\ref{susys}), it is easy to obtain two consistent condition
\bea \label{s}
\Gamma_{\star}\Gamma_{056}\epsilon_0=i\epsilon_0, ~~\Gamma_{01}K\epsilon_0=-\epsilon_0.
\eea
Thus the string solution preserves 1/4 of the supersymmetries.
  The total angular momentum carried by the string is given
\be J=\int P_{\phi}=\frac{L^2}{\pi
\alpha'}\int_0^{\sigma_{max}}d\sigma \sin^2\theta \dot{\phi}. \ee
After concrete calculation, the energy carried by the string is \be
E=\int P_t=\frac{L^2}{\pi \alpha'}\int_0^{\sigma_{max}} d\sigma
\cosh^2\rho \dot{t}=J. \ee These solutions satisfy the BPS
condition.

The string and some D3-brane solutions for 1/4 BPS Wilson loops can
be found from the supersymmetry conditions\cite{Drukker:2006zk}.
Following the same steps, we study the 1/4 BPS Wilson-'t Hooft loops
using D3-branes with both electric and magnetic charges on its
world-volume.

\subsection{D3-brane solution}

  The Wilson-'t Hooft loops have different representation of the gauge group.
In the bulk description, the existence of the D-string of the
Wilson-'t Hooft loop makes the general representation difficulty to
study. We will consider the case that the $F1$ strings and $D1$
stings are in the symmetric representation. The $F1$'s and $D1$'s
form a simple bound state.
  We will find a D3-brane solutions associated to this Wilson-'t Hooft loops.
The Wilson-'t Hooft loop that we consider is localized in the
time direction and preserves an $SO(3)\times SO(3)$ symmetry. These $SO(3)\times SO(3)$ are the isometry of $AdS_5$ and $S^5$ separately.
 It is convenient to use the metric (\ref{metric}) and fix a static gauge for D3-brane
with worldvolume coordinates $(t,\rho,\vartheta,\varphi)$. Then the D3-brane hypersurface in $AdS_5\times S^5$ is characterized by the equations
\bea \label{ansatz}
\chi=\chi(\rho),~~\theta=\theta(\rho),~~\phi=t.
\eea

  The D3-brane action includes three parts: the Dirac-Born-Infeld action (DBI), Wess-Zumio action (WZ) and boundary term
\be S=S_{DBI}+S_{WZ}+S_{boudary}=T_{D_3}\int
e^{-\Phi}\sqrt{-\det(g+2\pi\alpha'F)}-T_{D_3}\int
P[C_4]+S_{boudary}. \ee The tension of the D3-brane is \be
T_{D3}=\frac{N}{2\pi^2L^4}. \ee $P[C_4]$ is the pullback of the
Ramond-Ramond 4-form potential to the worldvolume of D3-brane. With
the ansatz (\ref{ansatz}), the DBI action is of the form(absorb a
factor of $\frac{2\pi\alpha'}{L^2}$ in the definition of
$F_{t\rho},F_{\vartheta\varphi}$) \be S_{DBI}=\frac{N}{2\pi^2}\int
dtd\rho d\vartheta
d\varphi\sqrt{(\cosh^2\rho-\sin^2\theta)(1+\sinh^2\rho
\chi'^2+\theta'^2)-F^2_{t\rho}}\sqrt{\sin^4\chi \sinh^4\rho
\sin^2\vartheta+F^2_{\vartheta\varphi}}. \ee Where $F_{t\rho}$ is
the elecric field and $F_{\vartheta \varphi}$ is the magnetic field
on D3-brane. The $'$ denotes a derivative with respect to $\rho$.
The Wess-Zumino term is given by \be S_{WZ}=\frac{2N}{\pi} \int
dtd\rho \sinh^4\rho \sin^2\chi \chi'. \ee It is difficult to solve
the equations of motion from this action. We will discuss the
supersymmetry equations of the D3-brane from requiring kappa
symmetry. These are first-order equations and can be solved easily.
These solutions can be checked to satisfy the equations of motion
derived from the action.
  The $\kappa$ symmetry of D-branes is defined as
  \cite{Bergshoeff:1996tu,Bergshoeff:1997kr,Skenderis:2002vf,Imamura:1998gk,Cederwall:1996pv,Aganagic:1996pe,Cederwall:1996ri,
Aganagic:1996nn}

\bea \label{kappa}
d^{p+1}\xi\Gamma&=&-e^{-\Phi}\mathcal{L}^{-1}_{DBI}e^{\mathcal{F}}\wedge X|_{vol}, \\
X&=&\bigoplus \Gamma_{(2n)}K^nI,
\eea

where the operations $I$ and $K$ act on spinors as $I\psi=-i\psi$
and $K\psi=\psi^{\ast}$, and the notation $\Gamma_{(n)}$ is defined
as \be \Gamma_{(n)}=\frac{1}{n!}d\xi^{i_n}\wedge\ldots\wedge
d\xi^{i_1}\Gamma_{i_1\cdots i_{n}}. \ee Where $\Gamma_{i_1\cdots
i_{n}}$ is the pullback of the target space gamma matrice \be
\Gamma_{i_1\cdots
i_{n}}=\partial_{i_1}X^{m_1}\ldots\partial_{i_n}X^{m_n}\Gamma_{m_1\cdots
m_n}. \ee The $\kappa$ symmetry of D3-brane with electric and
magnetic field on its world-volume is \bea
\Gamma=\mathcal{L}^{-1}_{DBI}(\Gamma_{(4)}+L^2F_{t\rho}\Gamma_{(2)}K+L^2F_{\vartheta\varphi}\tilde{\Gamma}_{(2)}K+L^4F_{t\rho}F_{\vartheta\varphi})I.
\eea The projectors $\Gamma_{(4)},\Gamma_{(2)}$ and
$\tilde{\Gamma}_{(2)}$ are \bea
\Gamma_{(4)}&=&\tilde{\Gamma}_{(2)}\Gamma_{(2)}=\partial_t x^{\mu}
\partial_{\rho} x^{\nu} \partial_{\vartheta}
x^{\xi}\partial_{\varphi}x^{\varsigma}\gamma_{\mu}
\gamma_{\nu}\gamma_{\xi}\gamma_{\zeta} =(\gamma_{t}+\gamma_{\phi})(\gamma_{\rho}+\chi'\gamma_{\chi}+\theta'\gamma_{\theta})\gamma_{\vartheta}\gamma_{\varphi}\\
\Gamma_{(2)}&=&\partial_{\vartheta}
x^{\mu}\partial_{\varphi}x^{\nu}\gamma_{\mu} \gamma_{\nu}
=\gamma_{\vartheta}\gamma_{\varphi}, ~~
\tilde{\Gamma}_{(2)}=(\gamma_{t}+\gamma_{\phi})(\gamma_{\rho}+\chi'\gamma_{\chi}+\theta'\gamma_{\theta}).
\eea

Using the Vielbeins
\bea
e^0=L\cosh\rho dt, ~~ e^1&=&L d\rho, ~~e^2=L\sinh\rho d\chi \\
e^3=L\sinh \rho \sin \chi d\vartheta, &~~&e^4=L\sinh \rho \sin \chi \sin\vartheta d\varphi \\
e^5=Ld\theta, &~~&e^6=L\sin\theta d\phi \eea
 and the ansatz (\ref{ansatz}),
the projectors $\Gamma_{(4)},\Gamma_{(2)}$ and
$\tilde{\Gamma}_{(2)}$ are written as \bea
\Gamma_{(4)}&=&\tilde{\Gamma}_{(2)}\Gamma_{(2)}=L^4(\cosh\rho\Gamma_0+\sin\theta\Gamma_6)(\Gamma_1+\sinh\rho\chi'\Gamma_2+
\theta'\Gamma_5)\sinh^2\rho \sin^2\chi \sin\vartheta\Gamma_{34},\\
\Gamma_{(2)}&=&L^2\sinh^2\rho \sin^2\chi \sin\vartheta\Gamma_{34},
~~
\tilde{\Gamma}_{(2)}=L^2(\cosh\rho\Gamma_0+\sin\theta\Gamma_6)(\Gamma_1+\sinh\rho\chi'\Gamma_2+\theta'\Gamma_5).
\eea The Killing spinor for the metric (\ref{metric}) is \be
\label{killings}
\epsilon=e^{-\frac{i}{2}\rho\Gamma_{\ast}\Gamma_1}e^{-\frac{i}{2}t\Gamma_{\ast}\Gamma_0}e^{\frac{1}{2}\chi\Gamma_{12}}e^{\frac{1}{2}\vartheta
F_{23}}e
^{\frac{1}{2}\varphi\Gamma_{34}}e^{-\frac{i}{2}\theta\Gamma_{\ast}\Gamma_5}e^{\frac{1}{2}\phi
F_{56}}\epsilon_0. \ee
 Similar to the Wilson loop case, we make the following ansatz on $\epsilon_0$ for 1/4 BPS Wilson-'t Hooft loops :
\be \label{cons}
K\epsilon_0=-e^{i\alpha}\Gamma_{01}\epsilon_0,~~\Gamma_6\epsilon_0=-i\Gamma_{12345}\epsilon_0.
\ee These conditions are different with the conditions (\ref{s}) of
Wilson loop with extra factor $e^{i\alpha}$. The $\alpha$ is the
real number which is similar to the angle. When $\alpha=0$, this
conditions reduce to the 1/4 BPS Wilson loop conditions (\ref{s})
corresponding the pure electric case.

  Putting the $\phi=t$ in the expression (\ref{killings}), the Killing spinors can be rewritten as
\be
\epsilon=e^{-\frac{i}{2}\rho\Gamma_{\ast}\Gamma_1}e^{-\frac{i}{2}\theta\Gamma_{\ast}\Gamma_5}e^{\frac{1}{2}\chi\Gamma_{12}}e^{\frac{1}{2}\vartheta
F_{23}}e
^{\frac{1}{2}\varphi\Gamma_{34}}\epsilon_0=e^{-\frac{i}{2}\rho\Gamma_{\ast}\Gamma_1}e^{-\frac{i}{2}\theta\Gamma_{\ast}\Gamma_5}e^{\frac{1}{2}\chi\Gamma_{12}}
M\epsilon_0. \ee Where \be M=e^{\frac{1}{2}\vartheta F_{23}}e
^{\frac{1}{2}\varphi\Gamma_{34}}. \ee We note that the $\epsilon_0$
and $M\epsilon_0$ satisfy the same constraints.

The differential equations for the 1/4 BPS Wilson-'t Hooft loops
come from the projection relation \be \Gamma\epsilon=\epsilon. \ee
After moving the matrix
$e^{-\frac{i}{2}\rho\Gamma_{\ast}\Gamma_1}e^{-\frac{i}{2}\theta\Gamma_{\ast}\Gamma_5}e^{\frac{1}{2}\chi\Gamma_{12}}$
to the left of the projector and applying the
constrains(\ref{cons}), we obtain the 8 differential equations for
the $\theta, \chi, F_{t\rho}$ and $F_{\vartheta\varphi}$

\bea &\Gamma_{0345}&:0=\sinh^2\rho \sin^2\chi
\sin\vartheta(e^{i\alpha}F_{t\rho}\sinh\rho \cos\chi
\sin\theta-\theta'(\cosh^2\rho-\sin^2\theta))-i
e^{i\alpha}F_{\vartheta\varphi}
                \sinh^3\rho \cos\chi \sin\theta \chi'     \nn  \\
&\Gamma_{\ast}\Gamma_5&:0=\sinh^2\rho \sin^2\chi
\sin\vartheta(e^{i\alpha}F_{t\rho} \sinh\rho \sin\chi
\sin\theta-\chi'\sinh^2\rho \sin\theta \cos\theta)-i
e^{i\alpha}F_{\vartheta\varphi}\sinh^3\rho
                        \sin\chi \sin\theta \chi' \nn \\
&\Gamma_{0234}&:0=(\cosh^2\rho-\sin^2\theta)\sin\chi+\chi'\cosh\rho \sinh\rho \cos\chi \cos^2\theta \nn \\
&\Gamma_{12}&:0=\sinh^2\rho \sin^2\chi \sin\vartheta(e^{i\alpha}F_{t\rho}\sinh\rho \cos\chi \cos\theta+\theta'\sin\theta \cos\theta+\cosh\rho \sinh\rho)\nn \\
&~~~~~~~~~~~~~&           -i
e^{i\alpha}F_{\vartheta\varphi}\sinh^3\rho
                       \cos\chi \cos\theta \chi' \nn
\eea
\bea
&\Gamma_{15}&:0=\chi'\cosh\rho \sinh\rho \cos\chi \sin\theta \cos\theta-\theta'\cosh\rho \sinh\rho \sin\chi+\sin\chi \sin\theta \cos\theta \nn \\
&\Gamma_{25}&:0=\sinh^2\rho \sin^2\chi \sin\vartheta(e^{i\alpha}F_{t\rho}\cosh\rho \sin\theta+\cos\chi \sin\theta \cos\theta-\chi'\cosh\rho \sinh\rho \sin\chi
\sin\theta \cos\theta- \nn \\
&~~~~~~~~~~~~~&                \theta'\cosh\rho \sinh\rho \cos\chi)-i e^{i\alpha}F_{\vartheta\varphi}\sinh^2\rho \cosh \rho  \sin\theta \chi' \nn \\
&\Gamma_{0134}&:0=\sinh^2\rho \sin^2\chi \sin\vartheta(e^{i\alpha}F_{t\rho}\cosh\rho \cos\theta+(\cosh^2\rho-\sin^2\theta)\cos\chi-\chi'\cosh\rho \sinh\rho
\sin\chi \cos^2\theta) \nn \\
&~~~~~~~~~~~~~& -i e^{i\alpha}F_{\vartheta\varphi}\sinh^2\rho \cosh\rho  \cos\theta \chi' \nn \\
&1&:-(\frac{1}{L^4})\mathcal{L}_{DBI}=\sinh^2\rho \sin^2\chi \sin\vartheta(e^{i\alpha}F_{t\rho}\sinh\rho \sin\chi \cos\theta+\chi'\sinh^2\rho \sin^2\theta)\nn \\
&~~~~~~~~~~~~~& -i e^{i\alpha}F_{\vartheta\varphi}\sinh^3\rho
\sin\chi \cos\theta \chi' -i
e^{i\alpha}F_{\vartheta\varphi}\frac{cos\theta
\sinh\rho}{\sin\chi}\chi'+i F_{t\rho}F_{\vartheta\varphi} \eea These
8 differential equations have 3 independent differential equations
 \bea \label{eq} \theta'&=&-\tan\theta \tanh\rho, \\
\label{eq1}
\chi'\cot\chi&=&-\frac{\cosh^2\rho-\sin^2\theta}{\cosh\rho \sinh\rho
\cos^2\theta}, \\ \label{eq2} e^{i\alpha}F_{t\rho}-i
e^{i\alpha}\chi' \sin^2\rho
\tilde{F}_{\vartheta\varphi}&=&-\frac{\cosh^2\rho-\sin^2\theta}{\cosh\rho
\cos\theta \cos\chi}, \eea where \bea \label{eer}
\tilde{F}_{\vartheta\varphi}=\frac{F_{\vartheta\varphi}}{\sinh^2\rho
\sin^2\chi \sin\vartheta}. \eea We solve for
$\chi,~\theta,~F_{t\rho}$ and $\tilde{F}_{\vartheta\varphi}$ from
these equations. The solutions to equations (\ref{eq}) and
(\ref{eq1}) are
 \be
\sin\theta=\frac{C_1}{\cosh\rho},
~~\sin\chi=C_2\frac{\coth\rho}{\sqrt{\cosh^2\rho-C_1^2}} . \ee Where
the $C_1$ and $C_2$ are two constants. The $C_1$ is related to the
angular momentum carried by the D3-brane. The $C_2$ is related to
the electric and magnetic charge of D3-brane. The electric and
magnetic field $F_{t\rho}, \tilde{F}_{\vartheta\varphi}$ can be
solved from the equation (\ref{eq2}). We can obtain two equations
from this complex equation (\ref{eq2}) by making the real part and
imaginary part equal separately. Finally, we get the results \be
F_{t\rho}=-\cos\alpha (\frac{\cosh^2\rho-\sin^2\theta}{\cosh\rho
\cos\theta \cos\chi}),~~
\tilde{F}_{\vartheta\varphi}=\frac{\cos\theta}{\sinh\rho
\sin\chi}\sin\alpha. \ee When $\cos\alpha=0$, the 1/4 BPS Wilson-'t
Hooft loops become the magnetic 't Hooft loops.

\subsection{Conserved charges}
The are two constants $C_1,~~C_2$ in the solutions. The $C_1$ is
related to the angular momentum J around the $S^2$ in the $S^5$. The
$C_2$ is related to the electric and magnetic charge carried by the
D3-brane.

  The electric charge k is the conjugate momentum to the electric field
\be \label{ggg} k=\Pi=\frac{2\pi \alpha'}{L^2}T_{D3}\int d\vartheta
d\varphi \frac{\delta \mathcal{L}}{\delta
F_{t\rho}}=\frac{4N}{\sqrt{\lambda}}C_2\cos\alpha. \ee The integer
charge $k$ corresponds to the number of coincident F1 stings. We
also have the magnetic charge $m$ \be \label{ggg1}
m=\frac{1}{2\pi}\frac{L^2}{2\pi\alpha'}\int d\vartheta d\varphi
F_{\vartheta\varphi}=\frac{\sqrt{\lambda}}{\pi}C_2\sin\alpha. \ee
The integer magnetic charge $m$ is the number of D1-branes immersed
in the D3-branes. From equations (\ref{ggg}) and (\ref{ggg1}), we
obtain the k and $m$ satisfy the following relation \be
\frac{k^2\lambda}{16N^2}+\frac{(m\pi)^2}{\lambda}=C_2^2. \ee

Using the dual coupling constant
$\tilde{\lambda}=\frac{16\pi^2N^2}{\lambda}$, the $C_2$ can be
rewritten as \be
C_2^2=\frac{k^2\lambda}{(4N)^2}+\frac{m^2\tilde{\lambda}}{(4N)^2}.
\ee
 We then discuss the energy and angular momentum relation.
The angular momentum J is
\bea
J&=&2T_{D3}\int d\vartheta d\varphi d\rho \frac{\delta \mathcal{L}}{\delta\dot{\phi}} \nn \\
&=&-\frac{4N}{\pi}\int d\rho \frac{\sinh^2\rho \sin^2\chi
\sin^2\theta(1+\sinh^2\rho
\chi'^2+\theta'^2)\sqrt{1+\tilde{F}^2_{\vartheta\varphi}}}{\sqrt{(\cosh^2\rho-\sin^2\theta)(1+\sinh^2\rho
\chi'^2+\theta'^2)-F^2_{t\rho}}}. \nn \eea The energy contribution
from DBI action and the Wess-Zumino term is \bea
E_{D.W}&=&2T_{D3}\int d\vartheta d\varphi d\rho \frac{\delta (\mathcal{L}_{DBI}+\mathcal{L}_{WZ})}{\delta\dot{t}}\nn \\
&=&\frac{4N}{\pi}\int d\rho \frac{\sinh^2\rho \sin^2\chi
\cosh^2\rho(1+\sinh^2\rho
\chi'^2+\theta'^2)\sqrt{1+\tilde{F}^2_{\vartheta\varphi}}}{\sqrt{(\cosh^2\rho-\sin^2\theta)(1+\sinh^2\rho
\chi'^2+\theta'^2)-F^2_{t\rho}}}+\frac{4N}{\pi}\int d\rho \sinh^4
\rho \sin^2\chi \chi' \nn \eea The other energy term comes from the
the electric and magnetic field. That is \bea
E_{L.T.}&=&\frac{4N}{\pi}\int  d\rho (-\frac{\delta
\mathcal{L}_{DBI}}{F_{\vartheta\varphi}}F_{\vartheta\varphi}
+\frac{\delta \mathcal{L}_{DBI}}
{F_{t\rho}}F_{t\rho}) \nn \\
&=&-\frac{4N}{\pi}\int d\rho
\frac{\sinh^2\rho \sin^2\chi \tilde{F}^2_{\vartheta\varphi}\sqrt{(\cosh^2\rho-\sin^2\theta)(1+\sinh^2\rho \chi'^2+\theta'^2)-
F^2_{t\rho}}}{\sqrt{1+\tilde{F}^2_{\vartheta\varphi}}} \nn \\
&-&\frac{4N}{\pi}\int d\rho \frac{\sinh^2\rho \sin^2\chi
F^2_{t\rho}\sqrt{1+\tilde{F}^2_{\vartheta\varphi}}}{\sqrt{(\cosh^2\rho-\sin^2\theta)(1+\sinh^2\rho
\chi'^2+\theta'^2)-F^2_{t\rho}}} \nn \eea Using the above
expressions, it is easy to find $J+E_{D.W}+E_{L.T.}=0$. This is
consistent with the BPS condition.

\section{1/4 BPS Circular Wilson-'t Hooft loop  }
\subsection{Review of the 1/4 BPS circular Wilson loop}
  In this section, we will discuss 1/4 BPS circular Wilson-'t Hooft loop. We first give a briefly review of the BPS circular Wilson loop
  \cite{Drukker:2005cu,Zarembo:2002an,Drukker:2006ga}. The 1/4 BPS circular Wilson loop
can be parameterized by \be x^1=R\cos\alpha,~~x^2=R \sin\alpha \ee
on the boundary of $AdS_5$. It couples to a linear combination
scalar \be \Phi(\alpha)=\Phi_3
\cos\theta_0+\sin\theta_0(\Phi_1\cos\alpha+\Phi_2 \sin\alpha) \ee
with an arbitrary fixed $\theta_0$. The loop may be written as (in
Euclidean signature) \be \label{kk}
W_{\theta_0}=Tr\mathcal{P}\exp[\oint(iA_{\mu}(\alpha)\dot{x}^{\mu}+|\dot{x}|\Phi(\alpha))d\alpha].
\ee The Wilson loop will be given by the sum of all non-interacting
diagrams which is written in terms of a 0-dimensional Hermitian
Gaussian matrix model. The expectation value of this Wilson loop is
given by the matrix model as \be \langle
W_{\theta_0}\rangle=\frac{1}{Z}\int
\mathcal{D}M\frac{1}{N}Tre^Me^{-\frac{2N}{\lambda'}TrM^2}=\frac{1}{N}L^1_{N-1}(-\frac{\lambda'}{4N})\exp[\frac{\lambda'}{8N}].
\ee Where $L^1_{N-1}$ is a Laguerre polynomial and $\lambda'=\lambda
\cos^2\theta_0$. In planar limit, the expectation value is given by
\be \langle
W_{\theta_0,planar}\rangle=\frac{2}{\sqrt{\lambda'}}I_1(\sqrt{\lambda'}).
\ee Where $I_1$ is a modified Bessel function.

To discuss the relevant string solutions in the dual $AdS_5\times
S^5$ space, it is useful to adopt the following metric on
$AdS_5\times S^2$ (the other directions on $S^5$ are omitted) \be
\label{cm1} \frac{ds^2}{L^2}=-d\chi^2+\cos^2\chi(d\rho^2+\sinh^2\rho
d\psi^2)+\sin^2\chi(d\sigma^2+\sinh^2\sigma
d\varphi^2)+d\theta^2+\sin^2\theta d\phi^2. \ee The metric has
Lorentzian signature. The Lorentzian can be used for the
supersymmetry analysis.

  The string describing the Wilson loop (\ref{kk}) will be at $\chi=0$ and end at $\rho\rightarrow \infty$. Using the ansatz
\be \rho=\rho(\sigma),~~ \psi(\tau)=\tau,~~
\theta=\theta(\tau),~~\phi(\tau)=\tau,~~\chi=0, \ee the string
action is \be \mathcal{S}=\frac{L^2}{4\pi\alpha'}\int d\sigma d\tau
[\rho'^2+\sinh^2\rho+\theta'^2+\sin^2\theta]. \ee The equations of
motion are \bea
\rho''&=&\sinh\rho \cosh\rho, \\
\theta''&=&\sin\theta \cos\theta. \eea Two solutions with these
boundary conditions were found\cite{Drukker:2006ga}: \be
\phi=\psi,~~\sinh\rho(\sigma)=\frac{1}{\sinh\sigma},~~\sin\theta=\frac{1}{\cosh(\sigma_0\pm\sigma)}.
\ee Here $\sigma$ is a world-sheet coordinate. The constant
$\sigma_0$ is fixed by the boundary condition that at $\sigma=0$ \be
\cos\theta_0=\tanh\sigma_0. \ee From these solutions, the classical
action is \be \mathcal{S}=\mp \cos\theta_0\sqrt{\lambda}. \ee The
two signs correspond to a string extended over the north or south
pole of $S^2$. The solutions preserve 1/4 of the supersymmetries.
This can be seen from supersymmetry analysis. The Killing spinor on
the relevant component of the metric (\ref{cm1}) is \be
\epsilon=e^{-\frac{i}{2}\rho
\Gamma_{\star}\Gamma_1}e^{\frac{1}{2}\psi
\Gamma_{12}}e^{-\frac{i}{2}\theta
\Gamma_{\star}\Gamma_5}e^{\frac{1}{2}\phi \Gamma_{56}}\epsilon_0.
\ee Then it can obtain two compatible constraints. \be
(\Gamma_{12}+\Gamma_{56})\epsilon_0=0 \ee and \be \label{sssss}
K\epsilon_0=-(\cos\theta_0\Gamma_{12}+\sin\theta_0\Gamma_{16})\epsilon_0.
\ee

The construction of the 1/4 BPS D3-brane describing the circular
Wilson loop in the k-th symmetric representation can be found in
\cite{Drukker:2006zk}. For k-th symmetric representation, the
expectation value is \be \langle
W_{k'}\rangle=\exp[2N(k'\sqrt{1+k'^2}+\arcsinh k')] \ee with \be
k'=\frac{k\cos\theta_0\sqrt{\lambda}}{4N}. \ee The $\theta_0=0$
corresponds to the 1/2 BPS Wilson loop (\ref{cj}).

We will turn on the magnetic flux on the D3-brane worldvolume to
find a 1/4 BPS circular Wilson-'t Hooft loop with D3-brane
description in the symmetric representation.
\subsection{D3-brane solution }
  We use the Lorentzian signature metric (\ref{cm1}) to obtain the equations of motion from the supersymmetry analysis.
  In this Lorentzian signature metric,
   the brane has extra factor of $i$ in the projector equation and electric field. We parameterize the
  D3 brane world-volume by $\{\rho,\psi,\sigma,\varphi\}$. The ansatz for the 1/4 BPS circular Wilson-'t Hooft loop
is to take $\chi=\chi(\rho), \theta=\theta(\rho), \psi=\phi$. The
killing spinor with Lorentzian signature metric (\ref{cm1}) can be
written as \be \label{kk2} \epsilon=e^{-\frac{i}{2}\chi
\Gamma_{\star}\Gamma_0}e^{-\frac{i}{2}\rho\Gamma_{\star}\Gamma_1}
e^{\frac{1}{2}\psi\Gamma_{12}}e^{-\frac{1}{2}\sigma\Gamma_{03}}e^{\frac{1}{2}\varphi\Gamma_{34}}
e^{-\frac{i}{2}\theta\Gamma_{\star}\Gamma_5}e^{\frac{1}{2}\phi\Gamma_{56}}\epsilon_0.
\ee The DBI action is of the form \be \mathcal{L}_{DBI}=L^4
\sin^2\chi \sinh\sigma
\sqrt{((-\chi'^2+\theta'^2+\cos^2\chi)(\cos^2\chi
\sinh^2\rho+\sin^2\theta)+F^2_{\rho\psi})(1-\tilde{F}^2_{\sigma\varphi})},
\ee where we define \be
\tilde{F}_{\sigma\varphi}=\frac{F_{\sigma\varphi}}{\sin^2\chi
\sinh\sigma}. \ee From the general $\kappa$ symmetry
expression(\ref{kappa}), we obtain \be
\Gamma=\mathcal{L}^{-1}_{DBI}(i\Gamma_{(4)}-L^2F_{\rho\psi}\Gamma_{(2)}K+iL^2F_{\sigma\varphi}\tilde{\Gamma}_{(2)}K-
L^4F_{\rho\psi}F_{\sigma\varphi})I. \ee The vielbeins for the metric
are \bea
e^0=Ld\chi,~~e^1&=&L\cos\chi d\rho,~~e^2=L \cos\chi \sinh \rho d\psi,\nn \\
e^3=L \sin\chi d\sigma,& ~~& e^4=L \sin\chi \sinh \sigma d\varphi, \nn \\
e^5=L d\theta, &~~& e^6=L \sin\theta d\phi. \eea Then the projectors
$\Gamma_{(4)}, \tilde{\Gamma}_{(2)}$ and $\Gamma_{(2)}$ can be
written as \bea
\Gamma_{(4)}&=&(\gamma_{\rho}+\chi'\gamma_{\chi}+\theta'\gamma_{\theta})(\gamma_{\psi}+\gamma_{\phi})
\gamma_{\sigma}\gamma_{\varphi} \\
            &=&L^4(\cos\chi\Gamma_1+\chi'\Gamma_0+\theta'\Gamma_5)(\cos\chi \sinh\rho\Gamma_2+\sin\theta\Gamma_6)\sin^2\chi \sinh\sigma\Gamma_{34},\\
\tilde{\Gamma}_{(2)}&=&(\gamma_{\rho}+\chi'\gamma_{\chi}+\theta'\gamma_{\theta})(\gamma_{\psi}+\gamma_{\phi})=
L^2(\cos\chi\Gamma_1+\chi'\Gamma_0+\theta'\Gamma_5)(\cos\chi \sinh\rho\Gamma_2+\sin\theta\Gamma_6),\\
\Gamma_{(2)}&=&\gamma_{\sigma}\gamma_{\varphi}=L^2\sin^2\chi
\sinh\sigma\Gamma_{34}. \eea The projector $\Gamma$ does not depend
on $\psi$. We can eliminate the dependence on $\psi$ by imposing \be
(\Gamma_{12}+\Gamma_{56})\epsilon_0=0. \ee After imposing this
constraint, the Killing spinor (\ref{kk2}) can be reduced to \be
\epsilon=e^{-\frac{i}{2}\chi
\Gamma_{\star}\Gamma_0}e^{-\frac{i}{2}\rho\Gamma_{\star}\Gamma_1}
e^{-\frac{1}{2}\sigma\Gamma_{03}}e^{\frac{1}{2}\varphi\Gamma_{34}}
e^{-\frac{i}{2}\theta\Gamma_{\star}\Gamma_5}\epsilon_0=e^{-\frac{i}{2}\chi
\Gamma_{\star}\Gamma_0}e^{-\frac{i}{2}\rho\Gamma_{\star}\Gamma_1}
e^{-\frac{i}{2}\theta\Gamma_{\star}\Gamma_5}M\epsilon_0. \ee Where
$M$ is defined as \be
M=e^{-\frac{1}{2}\sigma\Gamma_{03}}e^{\frac{1}{2}\varphi\Gamma_{34}}.
\ee To obtain the 1/4 BPS circular Wilson-'t Hooft loop, we impose
another condition \be
K\epsilon_0=-e^{i\gamma}(\cos\theta_0\Gamma_{12}+\sin\theta_0\Gamma_{16})\epsilon_0.
\ee This condition is different with the condition (\ref{sssss}) of
Wilson loop with extra factor $e^{i\gamma}$. The $\gamma$ is the
real number and connected with the electric and magnetic charge
carried by the D3-brane. The differential equations for the 1/4 BPS
Wilson-'t Hooft loops come from the projection relation \be
\Gamma\epsilon=\epsilon. \ee

Moving the matrix $e^{-\frac{i}{2}\chi
\Gamma_{\star}\Gamma_0}e^{-\frac{i}{2}\rho\Gamma_{\star}\Gamma_1}
e^{-\frac{i}{2}\theta\Gamma_{\star}\Gamma_5}$ to the left of
$\Gamma$, we obtain the 8 first order differential equations for
$\theta,\chi, F_{\rho\psi}$ and $\tilde{F}_{\sigma\varphi}$
\bea
&\Gamma_{0234}&: 0=i e^{i\gamma}F_{\rho\psi}\sin\chi \sin\theta
\sin\theta_0-ie^{i\gamma}\tilde{F}_{\sigma\varphi}(\sinh\rho
\sin\chi \cos\chi-\
\cosh\rho \chi')\sin\chi \sin\theta \sin\theta_0 \nn \\
&~~~~~~~~~~~~~&     +\chi'\sinh\rho(\cos^2\chi-\sin^2\theta)+\cosh\rho \sin\chi \cos\chi \sin^2\theta   \nn \\
&\Gamma_{\star}\Gamma_5&: 0=ie^{i\gamma} F_{\rho\psi}\sin\chi \sin\theta \cos\theta_0-ie^{i\gamma}\tilde{F}_{\sigma\varphi}(\sinh\rho \sin\chi \cos\chi-
\cosh\rho \chi')\sin\chi \sin\theta \cos\theta_0 \nn \\
&~~~~~~~~~~~~~&       -\chi'\cosh\rho \sin\theta \cos\theta+\sinh\rho \sin\chi \cos\chi \sin\theta \cos\theta  \nn \\
&\Gamma_{1234}&: 0=ie^{i\gamma} F_{\rho\psi} \sinh\rho \cos\chi \sin\theta \sin\theta_0 -ie^{i\gamma}\tilde{F}_{\sigma\varphi}(\sinh\rho \sin\chi \cos\chi-
\cosh\rho \chi')\sinh\rho \cos\chi \sin\theta \sin\theta_0 \nn \\
&~~~~~~~~~~~~~&  +i e^{i\gamma}F_{\rho\psi} \cosh\rho \cos\chi \cos\theta \cos\theta_0 -ie^{i\gamma}\tilde{F}_{\sigma\varphi}(\sinh\rho \sin\chi \cos\chi-
\cosh\rho \chi')\cosh\rho \cos\chi \cos\theta \cos\theta_0 \nn \\
&~~~~~~~~~~~~~& -\chi' \sinh^2\rho \sin\chi \cos\chi-\theta' \sin\theta \cos\theta +\sinh\rho \cosh\rho \cos^2\chi \nn \\
&\Gamma_{2345}& 0=ie^{i\gamma} F_{\rho\psi} \sinh\rho \cos\chi \sin\theta \cos\theta_0 -ie^{i\gamma}\tilde{F}_{\sigma\varphi}(\sinh\rho \sin\chi \cos\chi-
\cosh\rho \chi')\sinh\rho \cos\chi \sin\theta \cos\theta_0 \nn \\
&~~~~~~~~~~~~~&  -ie^{i\gamma} F_{\rho\psi} \cosh\rho \cos\chi \cos\theta \sin\theta_0 +ie^{i\gamma}\tilde{F}_{\sigma\varphi}(\sinh\rho \sin\chi \cos\chi-
\cosh\rho \chi')\cosh\rho \cos\chi \cos\theta \sin\theta_0 \nn \\
&~~~~~~~~~~~~~& -\theta'\sinh\rho \cosh\rho \cos^2\chi-\cos^2\chi \sin\theta \cos\theta \nn \\
&\Gamma_{01}& 0=ie^{i\gamma} F_{\rho\psi} \sinh\rho \cos\chi \cos\theta \cos\theta_0 -ie^{i\gamma}\tilde{F}_{\sigma\varphi}(\sinh\rho \sin\chi \cos\chi-
\cosh\rho \chi')\sinh\rho \cos\chi \cos\theta \cos\theta_0 \nn \\
&~~~~~~~~~~~~~&  +ie^{i\gamma} F_{\rho\psi} \cosh\rho \cos\chi \sin\theta \sin\theta_0 -ie^{i\gamma}\tilde{F}_{\sigma\varphi}(\sinh\rho \sin\chi \cos\chi-
\cosh\rho \chi')\cosh\rho \cos\chi \sin\theta \sin\theta_0 \nn \\
&~~~~~~~~~~~~~&-\chi'\sinh\rho \cosh\rho \sin\chi \cos\chi+\cos^2\chi(\sinh^2\rho+\sin^2\theta) \nn \\
&\Gamma_{05}&:0=i e^{i\gamma}F_{\rho\psi} \sinh\rho \cos\chi \cos\theta \sin\theta_0 -ie^{i\gamma}\tilde{F}_{\sigma\varphi}(\sinh\rho \sin\chi
\cos\chi-\cosh\rho \chi')\sinh\rho \cos\chi \cos\theta \sin\theta_0 \nn \\
&~~~~~~~~~~~~~&  -ie^{i\gamma} F_{\rho\psi} \cosh\rho \cos\chi \sin\theta \cos\theta_0 +ie^{i\gamma}\tilde{F}_{\sigma\varphi}(\sinh\rho \sin\chi
\cos\chi-\cosh\rho \chi')\cosh\rho \cos\chi \sin\theta \cos\theta_0 \nn \\
&~~~~~~~~~~~~~&+\theta'(\sinh^2\rho \cos^2\chi+\sin^2\theta) \nn \\
&\Gamma_{15}&: 0=ie^{i\gamma} F_{\rho\psi}\sin\chi \cos\theta \sin\theta_0-ie^{i\gamma}\tilde{F}_{\sigma\varphi}(\sinh\rho \sin\chi \cos\chi-
\cosh\rho \chi')\sin\chi \cos\theta \sin\theta_0 \nn \\
&~~~~~~~~~~~~~&-\chi' \sinh\rho \sin\theta \cos\theta+\theta'\sinh\rho \sin\chi \cos\chi+\cosh\rho \sin\chi \cos\chi \sin\theta \cos\theta \nn \\
&1&:1=-i L^4 \mathcal{L}^{-1}_{DBI}sin^2\chi \sinh\sigma(ie^{i\gamma} F_{\rho\psi}\sin\chi \cos\theta \cos\theta_0 -ie^{i\gamma}\tilde{F}_{\sigma\varphi}(\sin\chi
 \cos\theta \cos\theta_0 \nn \\
 &~~~~~~~~~~~~~&    -\frac{\cos\theta}{\sin\chi \cos \theta_0 })(\sinh\rho \sin\chi \cos\chi-\cosh\rho \chi')+\chi'\cosh\rho \sin^2\theta-
 \sinh\rho \sin\chi \cos\chi \sin^2\theta \nn \\
 &~~~~~~~~~~~~~&  -F_{\rho\psi}\tilde{F}_{\sigma\varphi}). \nn
\eea These equations are consistent with each other. We obtain three
independent equations \bea \label{jj} \theta'=A \cos^2\chi
\cos\theta&,& \chi'=A \sin\chi \cos\chi \sin\theta,\\ \label{334}
e^{i\gamma}F_{\rho\psi}-e^{i\gamma}\tilde{F}_{\sigma\varphi}(\sinh\rho
\sin\chi \cos\chi-\cosh\rho \chi')&=&-i\frac{\cos\chi
\cos\theta}{\cos\theta_0}(A\cosh\rho \sin\theta-\sinh\rho). \eea
Where \bea \label{F} A=\frac{\sinh\rho \cos\theta
\sin\theta_0-\cosh\rho \sin\theta
\cos\theta_0}{(\cos^2\chi-\sin^2\theta)\sinh\rho
\cos\theta_0+\cosh\rho \sin\theta \cos\theta \sin\theta_0 }. \eea
The electric field $F_{\rho\psi}$ and magnetic field
$\tilde{F}_{\sigma\varphi}$ can be solved from the equation
(\ref{334}) by taking the real part and imaginary part equal
separately. Then we then get the result \be
F_{\rho\psi}=-i\frac{\cos\chi \cos\theta}{\cos\theta_0}(A\cosh\rho
\sin\theta-\sinh\rho)\cos\gamma, ~~
\tilde{F}_{\sigma\varphi}=-\frac{\cos\theta}{\sin\chi
\cos\theta_0}\sin\gamma. \ee
  We can obtain the 1/4 BPS Wilson loop solution by setting $\gamma=0$. When $\cos\gamma$=0, these solutions are reduced to the 't Hooft loop.

From the equations (\ref{jj}), we can get the following results \bea
\label{v} \sin\chi \cos\theta&=&C, \\  \label{vv} \cos\chi(\cosh\rho
\cos\theta \sin\theta_0&-&\sinh\rho \sin\theta \cos\theta_0)=D. \eea
The solution not to be singular at the point $\rho=0$ require the D
satisfying the relation \be D=\pm \sin\theta_0\sqrt{1-C^2}, \ee
where the $+,-$ signs correspond to taking $\theta=0$ or
$\theta=\pi$ at $\rho=0$ respectively. These solutions in Lorentzian
space are unphysical. From these solutions we know that the
expectation value of the 1/4 BPS Wilson-'t Hooft loop for the fixed
$\theta_0$ depends on the $C,\gamma$ value.
\subsection{The 1/4 BPS Wilson-'t Hooft loop expectation value }

  We now discuss the expectation value of  1/4 BPS Wilson-'t Hooft loop $\langle
W_{\theta_0}^{WH}(C,\gamma)\rangle$. We don't know how to calculate
it from Super-Yang-Mills theory since we have to work in a
background with magnetic monopole. The $SL(2,Z)$ duality permits us
to get the answer. Thanks to the AdS/CFT duality, we can calculate
the expectation value of  1/4 BPS Wilson-'t Hooft loop using the
D3-brane action. We first point out that the expectation value is
independent of the $\gamma$ value. Then we use the 1/4 BPS Wilson
loop result to obtain the  expectation value of 1/4 BPS Wilson-'t
Hooft loop.
  \subsubsection{The expression of expectation value }

  The total on-shell D3-brane action
of 1/4 BPS Wilson-'t Hooft loop $S_{\theta_0}^{WH}(C,\gamma)$
includes four terms: the DBI action $S_{DBI}$, the Wess-Zumino
action $S_{WZ}$, the boundary term $S_F$ comes from the Legendre
transform of the gauge field and the other boundary term
$S_{\theta}$ comes from the Legendre transform of the scalar field
$\theta'$ \be
S_{\theta_0}^{WH}(C,\gamma)=S_{DBI}+S_{WZ}+S_F+S_{\theta}, \ee where
the DBI action is \bea
S_{DBI}&=&T_{D3}\int d\rho d\psi d\sigma d\varphi \mathcal{L}_{DBI} \nn \\
       &=&\frac{N}{2\pi^2}\int d\rho d\psi d\sigma d\varphi \sin^2\chi \sinh\sigma \sqrt{((-\chi'^2+\theta'^2+\cos^2\chi)
       (\cos^2\chi \sinh^2\rho+\sin^2\theta)+F^2_{\rho\psi})(1-\tilde{F}^2_{\sigma\varphi})} \nn \\
&=&\frac{-iN}{2\pi^2}\int d\rho d\psi d\sigma d\varphi \sin^3\chi
\sinh\sigma \cos\chi (A\cosh\rho \sin\theta-\sinh\rho)
(1-(\frac{\cos\theta}{\sin\chi \cos\theta_0})^2\sin^2\gamma) \eea
and the Legendre transform of the gauge field is \bea
S_F&=&-T_{D3}\int d\rho d\psi d\sigma d\varphi \frac{\delta\mathcal{L}_{DBI}}{\delta F_{\rho\psi}}F_{\rho\psi} \\
&=&\frac{-iN}{2\pi^2}\int d\rho d\psi d\sigma d\varphi \sin^3\chi
\sinh\sigma \cos\chi (A\cosh\rho \sin\theta-\sinh\rho)
(-(\frac{\cos\theta}{\sin\chi \cos\theta_0})^2\cos^2\gamma) \eea
Then \be S_{DBI}+S_F=\frac{-iN}{2\pi^2}\int d\rho d\psi d\sigma
d\varphi \sin^3\chi \sinh\sigma \cos\chi (A\cosh\rho
\sin\theta-\sinh\rho) (1-(\frac{\cos\theta}{\sin\chi
\cos\theta_0})^2) \ee is independent of $\gamma$. The $S_{WZ}$ term
also is independent of $\gamma$.

We finally only consider the $S_{\theta}$ term
\bea
S_{\theta}&=&-\theta_0'\int  d\psi d\sigma d\varphi P_{\theta}
=-T_{D3}\theta_0'\int d\psi d\sigma d\varphi \frac{\delta(\mathcal{L}_{DBI}+S_{WZ})}{\delta\theta'} \\
&=&-T_{D3}\theta_0'\int d\psi d\sigma d\varphi(\frac{i\sin\chi
\sinh\sigma \theta'(\cos^2\chi \sinh^2\rho+\sin^2\theta)}{\cos\chi
(A\cosh\rho \sin\theta-\sinh\rho)}+\frac{\delta
S_{WZ}}{\delta\theta'}). \eea Where $P_{\theta}$ is the conjugate
momentum of $\theta$. The $S_{\theta}$ is independent of $\gamma$
from above expression.

From the above analysis, we conclude that the total on-shell
D3-brane action of 1/4 BPS Wilson-'t Hooft loop is independent the
$\gamma$. This is to say that we can let the $\gamma=0$. As we know
that the $\gamma=0$ corresponds to the Wilson loop case. We have the
relation

\be \label{37}
S_{\theta_0}^{WH}(C,\gamma)=S_{\theta_0}^{WH}(C,0)=S_{\theta_0}^{Wilson
loop}(C). \ee  We can use the Wilson loop result to obtain the
Wilson-'t Hooft loop expectation value, but the $C$ has different
physical meaning.

The expectation value of 1/4 BPS Wilson loop has been calculated in
\cite{Drukker:2006zk}. The D3-brane solutions (\ref{v}), (\ref{vv})
and (\ref{F}) are unphysical in Lorentzian space. So we analytically
continue those solutions to Euclidean signature by taking the Wick
rotation \be \chi=iu,~~\sigma=i\vartheta. \ee

After this Wick rotation,  the Euclidean version of metric (\ref{cm1}) is

\be \frac{ds^2}{L^2}=du^2+\cosh^2u(d\rho^2+\sinh^2\rho
d\psi^2)+\sinh^2u(d\vartheta^2+\sin^2\vartheta
d\varphi^2)+d\theta^2+\sin^2\theta d\phi^2. \ee The solutions
(\ref{v}) and (\ref{vv}) become \bea
\sinh u\cos\theta&=&c, \\
\cosh u(\cosh\rho \cos\theta \sin\theta_0&-&\sinh\rho \sin\theta
\cos\theta_0)=d. \eea The solution smooth at $\rho=0$ only for \be
d=\pm \sin\theta_0\sqrt{1+c^2}. \ee From this, we can solved $\rho$
as a function $\theta$ \be
\sinh\rho=sign(\theta_0-\theta)\frac{\sin\theta
\sin\theta_0(\sqrt{1+c^2}\cos\theta_0+\cos\theta\sqrt{1+\frac{c^2\cos^2\theta_0}{\cos^2\theta}}}{\cosh
u(\cos^2\theta-\cos^2\theta_0)}. \ee Using this expression, the
solutions can be written as function of $\theta$ instead of $\rho$.
The world-volume is parameterized by
$\{\theta,\psi,\vartheta,\varphi\}$. The $\rho=\rho(\theta)$ and
$u=u(\theta)$ are given by the solutions above. The 1/4 BPS Wilson
loop has been discussed in \cite{Drukker:2006zk} in this
coordinates. The total D3-brane action of the 1/4 BPS Wilson loop is
given by \be S_{\theta_0}^{Wilson
loop}(c)=-2N(c\sqrt{1+c^2}+\arcsinh c). \ee

From the relation(\ref{37}), we find \be \label{355}
S_{\theta_0}^{WH}(c,\gamma)=-2N(c\sqrt{1+c^2}+\arcsinh c). \ee So
the expectation value of 1/4 BPS circular Wilson-'t Hooft loop is
\be \langle
W_{\theta_0}^{WH}(c,\gamma)\rangle=\exp[2N(c\sqrt{1+c^2}+\arcsinh
c)]. \ee

Although the expectation value of 1/4 BPS Circular Wilson-'t Hooft
loop has the same expression as Wilson loop, the $c$ has different
physical meaning.
\subsubsection{The physical meaning of $c$ }

The D3-brane DBI action with electric and magnetic field on its
world-volume is given by \be S_{DBI}^E=2N\int d\theta
d\vartheta\sqrt{(\cosh^2u
\rho'^2+u'^2+1)(\cosh^2u\sinh^2\rho+\sin^2\theta)+F^2_{\theta\psi}}\sqrt{\sinh^4u
\sin^2\vartheta+F^2_{\vartheta\varphi}}. \ee Where $'$ stands for
the derivative with respect to $\theta$. The momentum conjugate to
the gauge field $A_{\psi}$ \be
\Pi=-i\frac{2\pi\alpha'}{L^2}T_{D3}\int d\vartheta d\varphi
\frac{\delta\mathcal{L}_{DBI}^E}{\delta F_{\theta\psi}}=\pm
\frac{4N}{\sqrt{\lambda}}|\frac{c}{\cos\theta_0}|\cos\gamma=\pm k
.\ee The integer charge $k$ corresponds to the number of coincident
F1 stings.

The magnetic charge m is \be
m=\frac{1}{2\pi}\frac{L^2}{2\pi\alpha'}\int d\vartheta d\varphi
F_{\vartheta\varphi}=\frac{\sqrt{\lambda}}{\pi}\frac{c}{\cos\theta_0}\sin\gamma.
\ee The integer magnetic charge $m$ is the number of D1-branes
immersed in the D3-branes. From the above two expressions, we find
\be
c^2=\cos^2\theta_0(\frac{k^2\lambda}{16N^2}+\frac{(m\pi)^2}{\lambda}).
\ee  Using the dual coupling constant
$\tilde{\lambda}=\frac{16\pi^2N^2}{\lambda}$, $c$ can be written as
\be
c^2=\cos^2\theta_0[\frac{k^2\lambda}{(4N)^2}+\frac{m^2\tilde{\lambda}}{(4N)^2}].
\ee

Without considering the axion field, the expression  does not
manifest the $SL(2,Z)$ symmetry.
\subsubsection{The $SL(2,Z)$ symmetry}

To restore the full $SL(2,Z)$ duality, we must consider the effect
of a nonzero axion field $C_0$. For a non-zero constant axion
background, the D3-brane action has additional Wess-Zumino term \be
S_{axion}=\mu_3(2\pi \alpha)^2\int C_0
F_{\rho\psi}F_{\sigma\varphi}. \ee

For the nonvanishing axion field, the supersymmetry analysis still
holds because the kappa projection operator dosen't involve the any
RR filed. All the calculations are the same as the zero axion case
with a replacement \be k\rightarrow k+m C_0. \ee

From the AdS/CFT duality, the axion in the bulk could be identified
with the $\theta$ parameter in the Yang-Mills theory \be
C_0=\frac{\theta}{2\pi}. \ee

In the nonvanishing axion background $C_0$, the expectation value of
1/4 BPS circular Wilson-'t Hooft loop becomes \be \langle
W_{\theta_0}^{WH}(c,\gamma)\rangle=\exp[2N(c\sqrt{1+c^2}+\arcsinh
c)] \ee with \bea
c^2&=&\cos^2\theta_0[(k+\frac{m\theta}{2\pi})^2\frac{\lambda}{(4N)^2}+
\frac{m^2\tilde{\lambda}}{(4N)^2}] \nn \\
  &=&\cos^2\theta_0\frac{\pi}{4N}\frac{|k+m\tau|^2}{Im\tau}.
\eea
This is invariant under the $SL(2,Z)$ duality with $S$ and $T$ transformation
\bea
S~&:&~\tau \rightarrow -\frac{1}{\tau} ~~~~(n,m)\rightarrow(-m,n) \\
T~&:&~\tau\rightarrow \tau+1 ~~~~~~(n,m)\rightarrow(n+m,m). \eea

\section{Conclusions and Discussions }

  In this paper, we investigate the 1/4 BPS Wilson-'t Hooft loops in $\mathcal{N}$=4 supersymmetric
  Yang-Mills theory. We use the bulk D3-brane solutions with both electric and
magnetic charges to  describe some of 1/4 BPS Wilson-'t Hooft loops.
We consider 1/4 BPS Wilson-'t Hooft loops with both the $F1$'s and $D1$'s in symmetric representation.
Similar to the $F1$'s,
this can be studed by using the D3-branes configuration. We calculate the conserved charges for straight 1/4
BPS Wilson-'t Hooft loops and expectation value for circular 1/4 BPS Wilson-'t Hooft loops.

In our paper, we mainly discuss the 1/4 BPS Wilson-'t Hooft loops in
symmetry representation. It would be interesting to discuss the
general representation of 1/4 BPS Wilson-'t Hooft loops. Similar to
the case of Wilson loops, another interesting problem is the 1/4 BPS
Wilson surface in six-dimensional field theory in the framework of
$AdS_7/CFT_6$ correspondence\cite{Mal97,Seiberg97,Ganor:1996nf,
aharony,Nurmagambetov:2001ab,Chen,Chen:2007zzr,Aharony:2008ug,Chen:2008ds,Liu:2013uoa,Mori:2014tca}
using the $\kappa$ symmetry.

\section*{Acknowledgments }
 This work is supported by the NSF of China
Grant No. 11305131 and Research Start-up Foundation for Talents of
Northwest A\&F University of China Grant No. Z111021106, Z111021307.



\begin{thebibliography}{999}
\bibitem{Mal97}J.M.Maldacena, {\it ``The large $N$ limit of superconformal field
theories and supergravity,''}Adv.\ Theor.\ Math.\ Phys.\ {\bf 2},
231 (1998) [hep-th/9711200].

\bibitem{Gubser:1998bc}
  S.~S.~Gubser, I.~R.~Klebanov and A.~M.~Polyakov,
{\it  ``Gauge theory correlators from non-critical string theory,''}
  Phys.\ Lett.\  B {\bf 428}, 105 (1998)
  [arXiv:hep-th/9802109].

\bibitem{Witten:1998qj}
  E.~Witten,
{\it ``Anti-de Sitter space and holography,''}
  Adv.\ Theor.\ Math.\ Phys.\  {\bf 2}, 253 (1998)
  [arXiv:hep-th/9802150].\\

\bibitem{Aharony:1999ti}
  O.~Aharony, S.~S.~Gubser, J.~M.~Maldacena, H.~Ooguri and Y.~Oz,
{\it  ``Large N field theories, string theory and gravity,''}
  Phys.\ Rept.\  {\bf 323}, 183 (2000)
  [arXiv:hep-th/9905111].




\bibitem{Rey:1998ik}
  S.~J.~Rey and J.~T.~Yee,
 {\it ``Macroscopic strings as heavy quarks in large {N} gauge theory and  anti-de
  {S}itter supergravity,''}
  Eur.\ Phys.\ J.\  C {\bf 22}, 379 (2001)
  [arXiv:hep-th/9803001].

\bibitem{Mal98}J.~M. Maldacena, {\it ``{W}ilson loops in large {N} field
theories''},  Phys.
  Rev. Lett. {\bf 80} (1998) 4859--4862 [hep-th/9803002]\\

\bibitem{Gross}N.~Drukker, D.~J. Gross and H.~Ooguri, {\it ``{W}ilson loops and
minimal
  surfaces,''}  Phys. Rev. {\bf D60} (1999) 125006 [hep-th/9904191].

\bibitem{Zarembo}J.~K. Erickson, G.~W. Semenoff and K.~Zarembo, {\it ``Wilson
loops in {N = 4}
  supersymmetric {Y}ang-{M}ills theory,''}  Nucl. Phys. {\bf B582} (2000)
  155--175 [hep-th/0003055].


\bibitem{Drukker}
N.~Drukker and B.~Fiol, {\it ``All-genus calculation of {W}ilson
loops using
  {D}-branes,''}  JHEP {\bf 0502}, 010 (2005) [hep-th/0501109].

\bibitem{Yamaguchi:2006D5}
S.~Yamaguchi, {\it ``Wilson Loops of Anti-symmetric Representation
and D5-branes,''} JHEP {\bf 0605}, 037 (2006) [hep-th/0603208].

\bibitem{Gomis:2006sb}
  J.~Gomis and F.~Passerini,
{\it ``Holographic Wilson loops,''}
  JHEP {\bf 0608}, 074 (2006)
  [arXiv:hep-th/0604007].

\bibitem{Gomis:2006im}
  J.~Gomis and F.~Passerini,
{\it  ``Wilson loops as D3-branes,''}
  JHEP {\bf 0701}, 097 (2007)
  [arXiv:hep-th/0612022].



\bibitem{Drukker01} N.~Drukker and D.~J. Gross, {\it ``An exact prediction of {N = 4}
{SUSYM}
  theory for string theory,''}  J. Math. Phys. {\bf 42} (2001) 2896--2914 [hep-th/0010274].

\bibitem{Akemann:2001st}
  G.~Akemann and P.~H.~Damgaard,
  ``Wilson loops in N=4 supersymmetric Yang-Mills theory from random matrix theory,''
  Phys.\ Lett.\ B {\bf 513}, 179 (2001)
  [Phys.\ Lett.\ B {\bf 524}, 400 (2002)]
  [hep-th/0101225].

\bibitem{Myers:1999}
R.C. Myers, {\it ``Dielectric-Branes,''}  JHEP {\bf 9912}, 022
(1999) [hep-th/9910053].

\bibitem{Rodriguez2006}D. Rodriguez-Gomez, {\it ``Computing Wilson
lines with dielectric branes"}, Nucl. Phys. {\bf B752} (2006)
316-326 [hep-th/0604031].






\bibitem{Okuyama:2006jc}
  K.~Okuyama and G.~W.~Semenoff,
  {\it ``Wilson loops in N = 4 SYM and fermion droplets,''}
  JHEP {\bf 0606}, 057 (2006)
  [arXiv:hep-th/0604209].

\bibitem{Yamaguchi:2006te}
  S.~Yamaguchi,
  ``Bubbling geometries for half BPS Wilson lines,''
  Int.\ J.\ Mod.\ Phys.\ A {\bf 22}, 1353 (2007)
  [hep-th/0601089].
\bibitem{Lunin:2006xr}
  O.~Lunin,
 ``On gravitational description of Wilson lines,''
  JHEP {\bf 0606}, 026 (2006)
  [hep-th/0604133].
\bibitem{D'Hoker:2007fq}
  E.~D'Hoker, J.~Estes and M.~Gutperle,
 ``Gravity duals of half-BPS Wilson loops,''
  JHEP {\bf 0706}, 063 (2007)
  [arXiv:0705.1004 [hep-th]].

\bibitem{Hartnoll:2006is}
  S.~A.~Hartnoll and S.~P.~Kumar,
  {\it ``Higher rank Wilson loops from a matrix model,''}
  JHEP {\bf 0608}, 026 (2006)
  [arXiv:hep-th/0605027].
\bibitem{Faraggi:2014tna}
  A.~Faraggi, J.~T.~Liu, L.~A.~Pando Zayas and G.~Zhang,
  ``One-loop structure of higher rank Wilson loops in AdS/CFT,''
  Phys.\ Lett.\ B {\bf 740}, 218 (2015)
  [arXiv:1409.3187 [hep-th]].

\bibitem{Grisaru:2000zn}
  M.~T.~Grisaru, R.~C.~Myers and O.~Tafjord,
  ``SUSY and goliath,''
  JHEP {\bf 0008}, 040 (2000)
  [hep-th/0008015].

\bibitem{Hashimoto:2000zp}
  A.~Hashimoto, S.~Hirano and N.~Itzhaki,
  ``Large branes in AdS and their field theory dual,''
  JHEP {\bf 0008}, 051 (2000)
  [hep-th/0008016].

\bibitem{McGreevy:2000cw}
  J.~McGreevy, L.~Susskind and N.~Toumbas,
  ``Invasion of the giant gravitons from Anti-de Sitter space,''
  JHEP {\bf 0006}, 008 (2000)
  [hep-th/0003075].

\bibitem{Lin:2004nb}
  H.~Lin, O.~Lunin and J.~M.~Maldacena,
  `Bubbling AdS space and 1/2 BPS geometries,''
  JHEP {\bf 0410}, 025 (2004)
  [hep-th/0409174].



\bibitem{Kapustin:2005py}
  A.~Kapustin,
  ``Wilson-'t Hooft operators in four-dimensional gauge theories and S-duality,''
  Phys.\ Rev.\ D {\bf 74}, 025005 (2006)
  [hep-th/0501015].

\bibitem{Pucci:2012sx}
  F.~Pucci,
  ``More on 't Hooft loops in N=4 SYM,''
  JHEP {\bf 1211}, 161 (2012)
  [arXiv:1207.6627 [hep-th]].


\bibitem{Witten:1995im}
  E.~Witten,
  ``Bound states of strings and p-branes,''
  Nucl.\ Phys.\ B {\bf 460}, 335 (1996)
  [hep-th/9510135].
\bibitem{ChenHe}B. Chen and W. He, {\it ``1/2 BPS Wilson-'t Hooft
loops"}, Phys. Rev. D{\bf 74}(2006)126008 [hep-th/0607024].

\bibitem{Drukker:2006xg}
  N.~Drukker and S.~Kawamoto,
  ``Small deformations of supersymmetric Wilson loops and open spin-chains,''
  JHEP {\bf 0607}, 024 (2006)
  [hep-th/0604124].

\bibitem{Drukker:2006zk}
  N.~Drukker, S.~Giombi, R.~Ricci and D.~Trancanelli,
{\it ``On the D3-brane description of some 1/4 BPS Wilson loops,''}
  JHEP {\bf 0704}, 008 (2007)
  [arXiv:hep-th/0612168].


\bibitem{Bergshoeff:1996tu}
  E.~Bergshoeff and P.~K.~Townsend,
  ``Super D-branes,''
  Nucl.\ Phys.\ B {\bf 490}, 145 (1997)
  [hep-th/9611173].


\bibitem{Bergshoeff:1997kr}
  E.~Bergshoeff, R.~Kallosh, T.~Ortin and G.~Papadopoulos,
  ``Kappa symmetry, supersymmetry and intersecting branes,''
  Nucl.\ Phys.\ B {\bf 502}, 149 (1997)
  [hep-th/9705040].

\bibitem{Skenderis:2002vf}
  K.~Skenderis and M.~Taylor,
  ``Branes in AdS and p p wave space-times,''
  JHEP {\bf 0206}, 025 (2002)
  [hep-th/0204054].

\bibitem{Imamura:1998gk}
  Y.~Imamura,
  ``Supersymmetries and BPS configurations on Anti-de Sitter space,''
  Nucl.\ Phys.\ B {\bf 537}, 184 (1999)
  [hep-th/9807179].
\bibitem{Cederwall:1996pv}
  M.~Cederwall, A.~von Gussich, B.~E.~W.~Nilsson and A.~Westerberg,
  ``The Dirichlet super three-brane in ten-dimensional type IIB supergravity,''
  Nucl.\ Phys.\ B {\bf 490}, 163 (1997)
  [hep-th/9610148].
\bibitem{Aganagic:1996pe}
  M.~Aganagic, C.~Popescu and J.~H.~Schwarz,
  ``D-brane actions with local kappa symmetry,''
  Phys.\ Lett.\ B {\bf 393}, 311 (1997)
  [hep-th/9610249].
\bibitem{Cederwall:1996ri}
  M.~Cederwall, A.~von Gussich, B.~E.~W.~Nilsson, P.~Sundell and A.~Westerberg,
  ``The Dirichlet super p-branes in ten-dimensional type IIA and IIB supergravity,''
  Nucl.\ Phys.\ B {\bf 490}, 179 (1997)
  [hep-th/9611159].
\bibitem{Aganagic:1996nn}
  M.~Aganagic, C.~Popescu and J.~H.~Schwarz,
  ``Gauge invariant and gauge fixed D-brane actions,''
  Nucl.\ Phys.\ B {\bf 495}, 99 (1997)
  [hep-th/9612080].


\bibitem{Drukker:2005cu}
  N.~Drukker and B.~Fiol,
  ``On the integrability of Wilson loops in AdS(5) x S**5: Some periodic ansatze,''
  JHEP {\bf 0601}, 056 (2006)
  [hep-th/0506058].
\bibitem{Zarembo:2002an}
  K.~Zarembo,
  ``Supersymmetric Wilson loops,''
  Nucl.\ Phys.\ B {\bf 643}, 157 (2002)
  [hep-th/0205160].

\bibitem{Drukker:2006ga}
  N.~Drukker,
  ``1/4 BPS circular loops, unstable world-sheet instantons and the matrix model,''
  JHEP {\bf 0609}, 004 (2006)
  [hep-th/0605151].



\bibitem{Seiberg97} O. Aharony, M. Berkooz, S. Kachru, N. Seiberg and E. Silverstein, {\it
``Matrix description of interacting theories in six-dimensions"},
Adv. Theor. Math. Phys. 1 (1998) 148-157,
[hep-th/9707079].\\

\bibitem{Ganor:1996nf}
  O.~J.~Ganor,
  ``Six-dimensional tensionless strings in the large N limit,''
  Nucl.\ Phys.\ B {\bf 489}, 95 (1997)
  [hep-th/9605201].

\bibitem{aharony}
 O. Aharony, M. Berkooz and N. Seiberg, {\it ``
Light-Cone Description of (2,0) Superconformal Theories in Six
Dimensions"}, Adv. Theor. Math. Phys. 2 (1998) 119-153,
[hep-th/9712117].







\bibitem{Nurmagambetov:2001ab}
  A.~J.~Nurmagambetov and I.~Y.~Park,
  {\it ``On the M5 and the AdS(7)/CFT(6) correspondence,''}
  Phys.\ Lett.\  B {\bf 524}, 185 (2002)
  [arXiv:hep-th/0110192].




\bibitem{Chen}
  B.~Chen, W.~He, J.~B.~Wu and L.~Zhang,
  {\it ``M5-branes and Wilson Surfaces,''}
  JHEP {\bf 0708}, 067 (2007)
  [arXiv:0707.3978 [hep-th]].




\bibitem{Chen:2007zzr}
  B.~Chen, C.~Y.~Liu and J.~B.~Wu,
  ``Operator Product Expansion of Wilson surfaces from M5-branes,''
  JHEP {\bf 0801}, 007 (2008)
  [arXiv:0711.2194 [hep-th]].



\bibitem{Aharony:2008ug}
  O.~Aharony, O.~Bergman, D.~L.~Jafferis and J.~Maldacena,
  ``N=6 superconformal Chern-Simons-matter theories, M2-branes and their gravity duals,''
  JHEP {\bf 0810}, 091 (2008)
  [arXiv:0806.1218 [hep-th]].

\bibitem{Chen:2008ds}
  B.~Chen and J.~B.~Wu,
  ``Wilson-Polyakov surfaces and M-theory branes,''
  JHEP {\bf 0805}, 046 (2008)
  [arXiv:0802.2173 [hep-th]].

\bibitem{Liu:2013uoa}
  C.~Y.~Liu,
  ``Wilson surface correlator in the $AdS_7/CFT_6$ correspondence,''
  JHEP {\bf 1307}, 009 (2013).
\bibitem{Mori:2014tca}
  H.~Mori and S.~Yamaguchi,
  ``M5-branes and Wilson Surfaces in AdS$_{7}$/CFT$_{6}$ Correspondence,''
  Phys.\ Rev.\ D {\bf 90}, 026005 (2014)
  [arXiv:1404.0930 [hep-th]].

\end{thebibliography}
\end{document}